\documentclass{eas}
\usepackage{graphicx}

\begin{document}

\TitreGlobal{Dark Energy and Dark Matter: Observations, Experiments and Theories}

\Editors{E. Pecontal, T. Buchert, Y. Copin, P. Di Stefano (eds)}

\title{Dark Energy vs. Dark Matter: Towards a Unifying Scalar Field?}
\author{Arbey, A.}\address{Universit\'e de Lyon, Lyon, F-69000, France ; Universit\'e Lyon~1, Villeurbanne, F-69622, France ; Centre de Recherche Astrophysique de Lyon, Observatoire de Lyon, 9 avenue Charles Andr\'e, Saint-Genis Laval cedex, F-69561, France ; CNRS, UMR 5574 ; Ecole Normale Sup\'erieure de Lyon, Lyon, France -- Email: arbey@obs.univ-lyon1.fr}
\runningtitle{Unifying scalar field dark fluid}
\begin{abstract}
The standard model of cosmology suggests the existence of two components, ``dark
matter'' and ``dark energy'', which determine the fate of the Universe. Their nature is still under investigation, and no direct proof of their existences has emerged yet. There exist alternative models which reinterpret the cosmological observations, for example by replacing the dark energy/dark matter hypothesis by the existence of a unique dark component, the {\it dark fluid}, which is able to mimic the behaviour of both components. After a quick review of the cosmological constraints on this unifying dark fluid, we will present a model of dark fluid based on a complex scalar field and discuss the problem of the choice of the potential.
\end{abstract}

\maketitle

\section{Introduction}
The standard model of cosmology suggests that the total energy density of the Universe is presently dominated by the densities of two components: the {\it dark matter} component, which is a pressureless matter fluid having an attractive behaviour, and the {\it dark energy}, whose main properties are a negative pressure and a nearly constant energy density today, which evoke the idea of vacuum energy. The nature of both components remains unknown, and in the near future we can hope that the Large Hadron Collider (LHC) will be able to provide hints on the nature of dark matter. In spite of the mysteries of the dark components, it is generally considered that dark matter can be modeled as a system of collisionless particles, whereas the most usual models of dark energy are the scalar field based quintessence models. However, many difficulties in usual dark energy and dark matter models still question the cosmological standard model bases, leaving room for other models to be investigated. In this paper, we consider a unifying model in which the dark matter and dark energy
components can be considered as two different aspects of a same component, the {\it dark fluid}. We will first review how such a unifying model can be constrained by the cosmological observations. Then we will consider a dark fluid model based on a complex scalar field, and we will discuss the question of the choice of the scalar field potential.

\section{Constraints on dark fluid models}
During last years, cosmological observations have greatly improved in precision and in number. Whereas their analyses are generally performed within the standard model of cosmology, considering distinctly dark matter and dark energy, they can be reinterpreted to determine constraints on the dark fluid model. A first analysis of the cosmological constraints has been performed in (Arbey, 2005). With new data becoming available, and in particular with the new 5-year WMAP constraints (Komatsu et al., 2008), a reanalysis is performed in (Arbey, 2008). We will present here some of the most interesting results.\\
First we define as usual $\Omega_D$ as the ratio of the density of the dark fluid over the critical density, and $\omega_D$ as the ratio of the pressure of the dark fluid over its density.\\
According to (Arbey, 2008) recent observations of supernov\ae~of type Ia impose the following constraints on the dark fluid:
\begin{equation}
\Omega_D^0 = 1.005 \pm 0.006 \;\;,\;\;\omega_D^0 = -0.80 \pm 0.12 \;\;,\;\;
\omega_D^a = 0.9 \pm 0.5 \;\;,
\end{equation}
where $\omega_D$ is written in function of the expansion factor:
\begin{equation}
\omega_D= \omega_D^0 + (1-a) \omega_D^a \;\;,
\end{equation}
with $a_0=1$. The result on $\omega_D^0$ is particularly interesting, as it reveals that $\omega_D^0 > -1$, which is a property that has to be satisfied to enable a description of the dark fluid by a scalar field.\\
\\
Structure formation is also a very interesting way to contrain cosmological models. However, no stringent constraints can be determined without performing a precise analysis of a specific dark fluid model. We have nevertheless shown that a dark fluid should have an equation of state respecting $\omega_D > -1/3$ at the time of structure formation.\\
\\
The data on the Cosmic Microwave Background (CMB) have been greatly improved with the new WMAP 5-year data. If only the CMB data are considered, a large variety of dark fluid models would still be permitted. We can however consider that a dark fluid model is more realistic if the fluid behaves today like a dark energy with a negative pressure, but was behaving mainly like matter at the recombination time. We refer to (Arbey, 2008) for a more thourough analysis of the CMB constraints.\\
\\
Constraints can also be derived considering primordial nucleosynthesis (BBN) models. Under the assumption that the Universe is dominated by radiation at BBN time, we expect the dark fluid density to be small in comparison to the radiation density, in order to prevent the expansion rate to be strongly modified at this period, since it would lead to a modification in the abundance of the elements. If we assume that the dark fluid behaviour does not change violently during BBN, it means that the equation of state of this fluid during the BBN period has to be $\omega_D(\mbox{BBN}) \leq 1/3$, or that its density was completely negligible before BBN~\\
\\
Taken separately, these constraints does not seem very severe, but in a complete dark fluid scenario, all of them must be respected simultaneously, and when requiring that the dark fluid should also be able to explain dark matter at local scales, they become very restrictive.\\
We will now consider a dark fluid model based on a complex scalar field and confront it with the observational constraints.

\section{Complex scalar field dark fluid model}
\begin{figure}[h]
\centering
\includegraphics[width=4.3cm,angle=270]{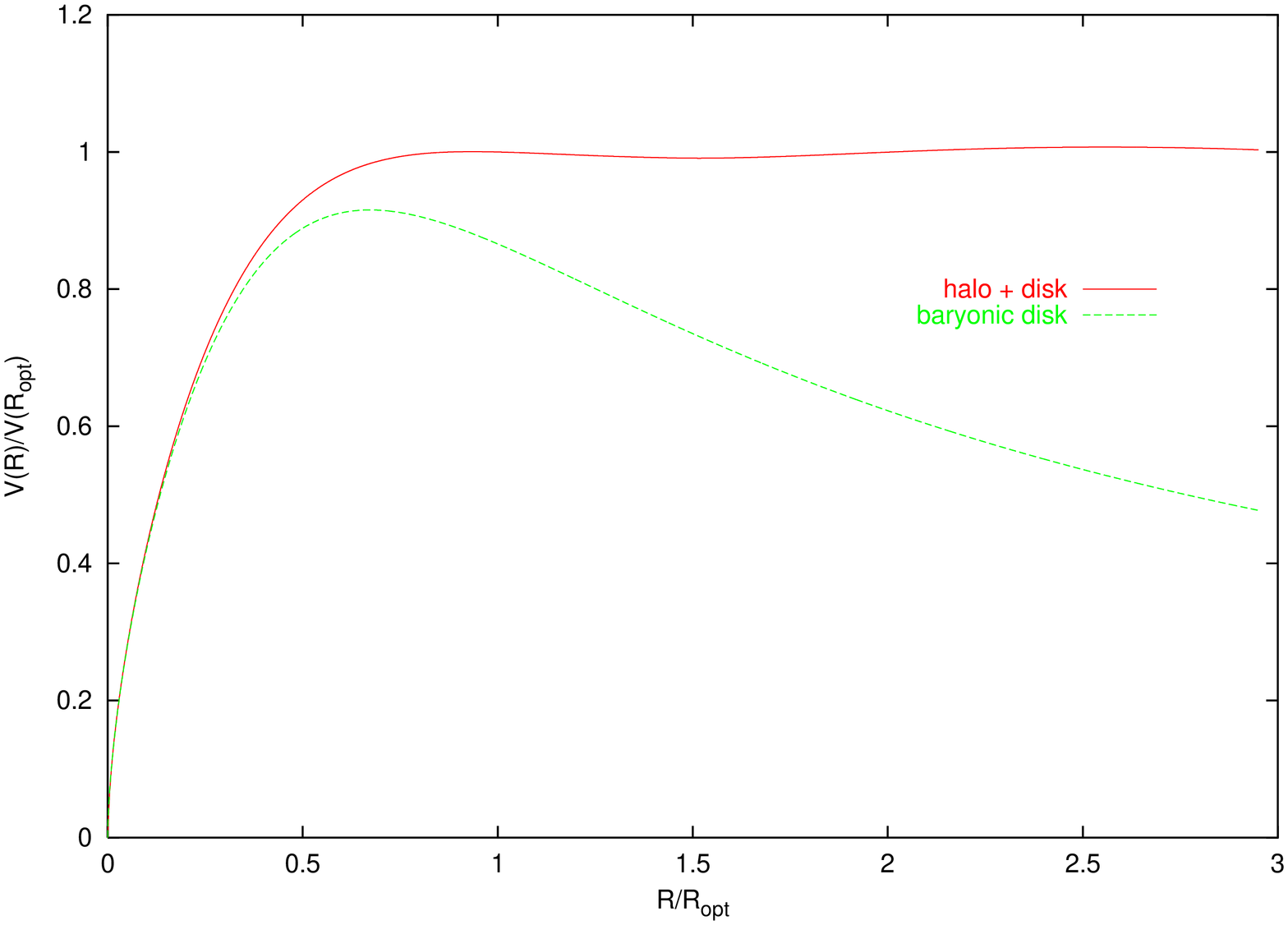}
\includegraphics[width=4.3cm,angle=270]{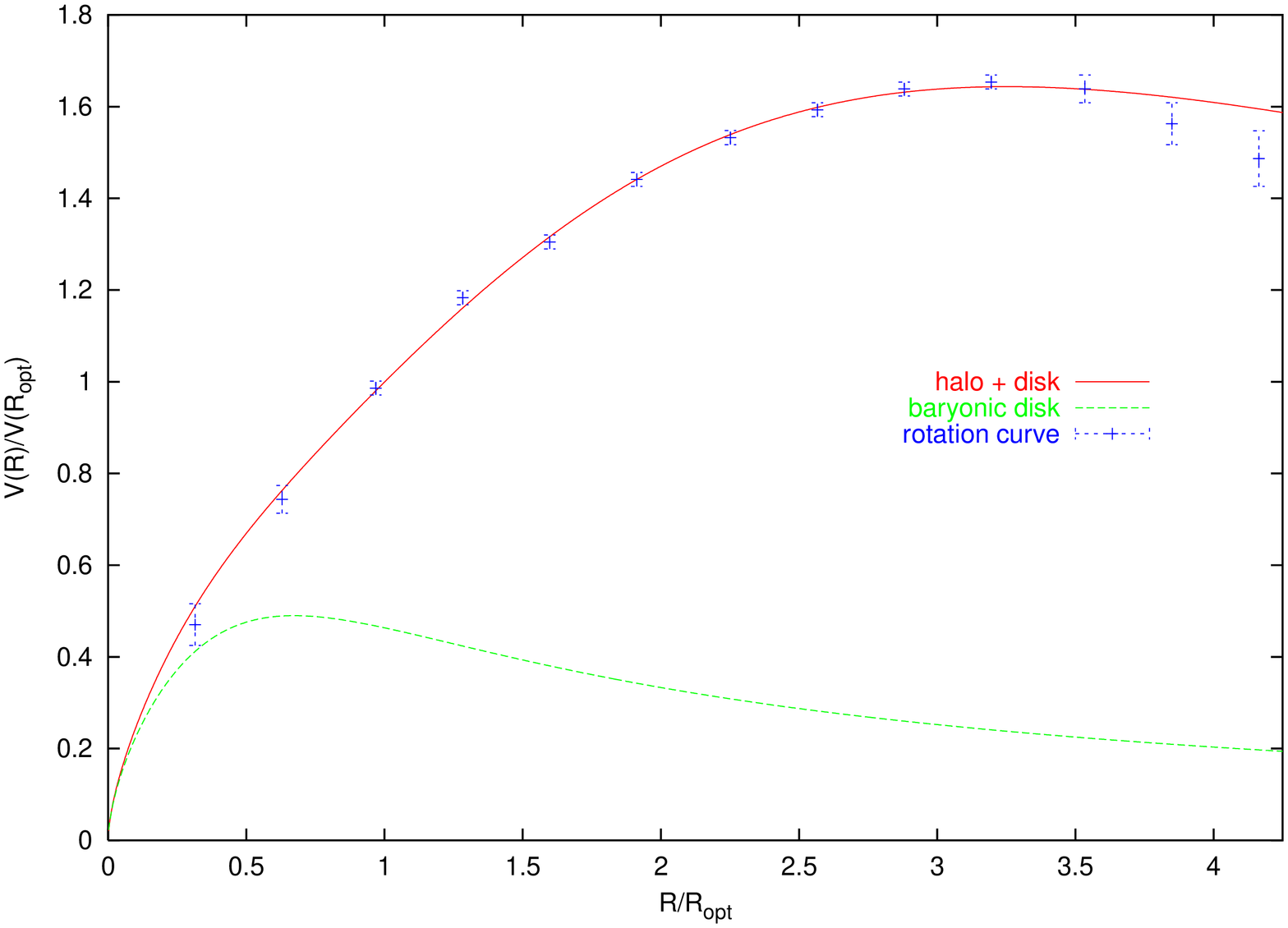}
\caption{On the left plot, flat galaxy rotation curve (solid line) induced by the
presence of a complex scalar field with a quadratic potential. The dashed line
corresponds to the contribution from baryonic matter only. On the right plot, rotation curve of the dwarf spiral DDO 154 reproduced by the same massive complex scalar field.}
\label{gal}
\end{figure}

\begin{figure}[h]
\centering
\includegraphics [width=4.4cm,angle=270] {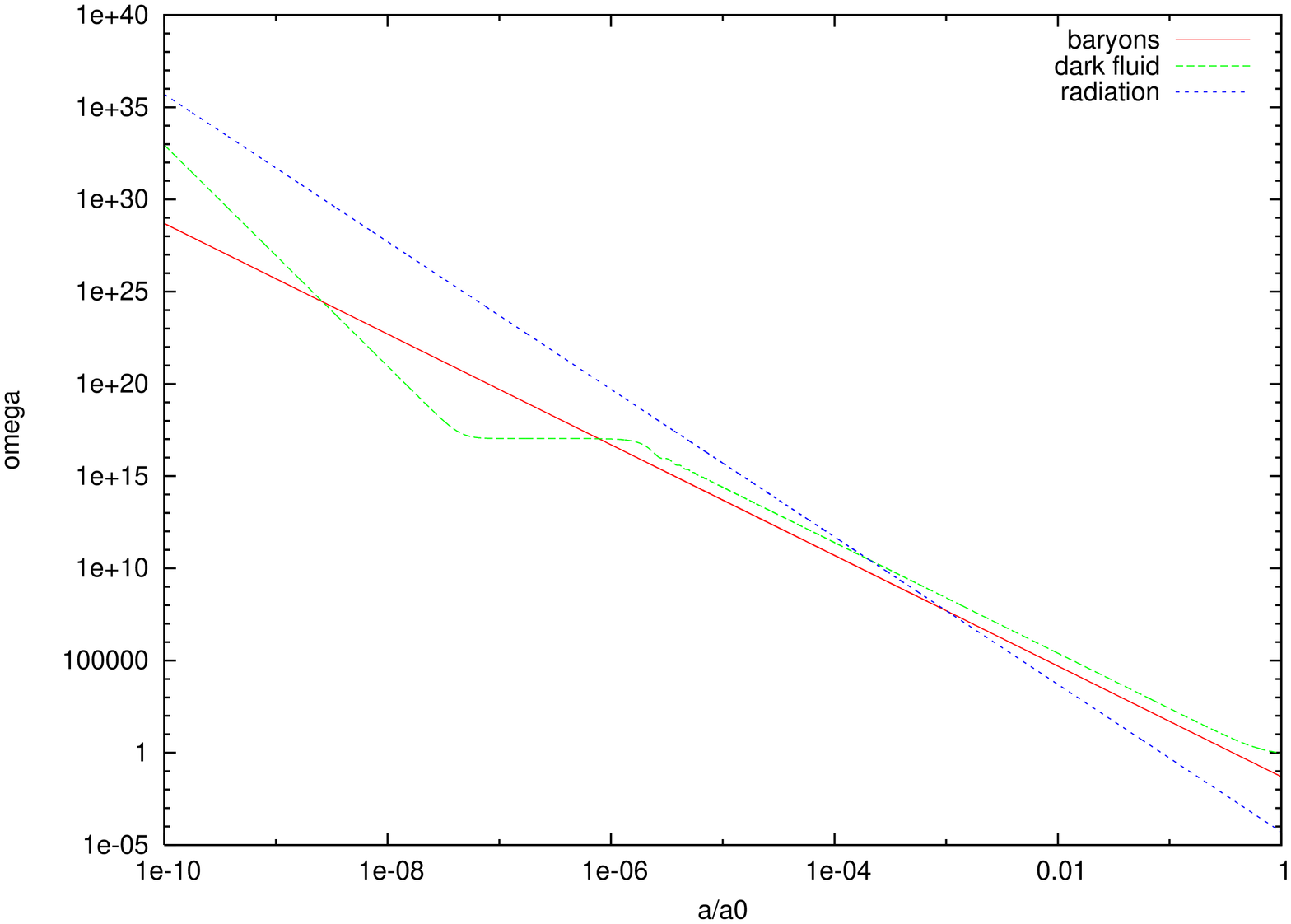}\includegraphics [width=4.4cm,angle=270] {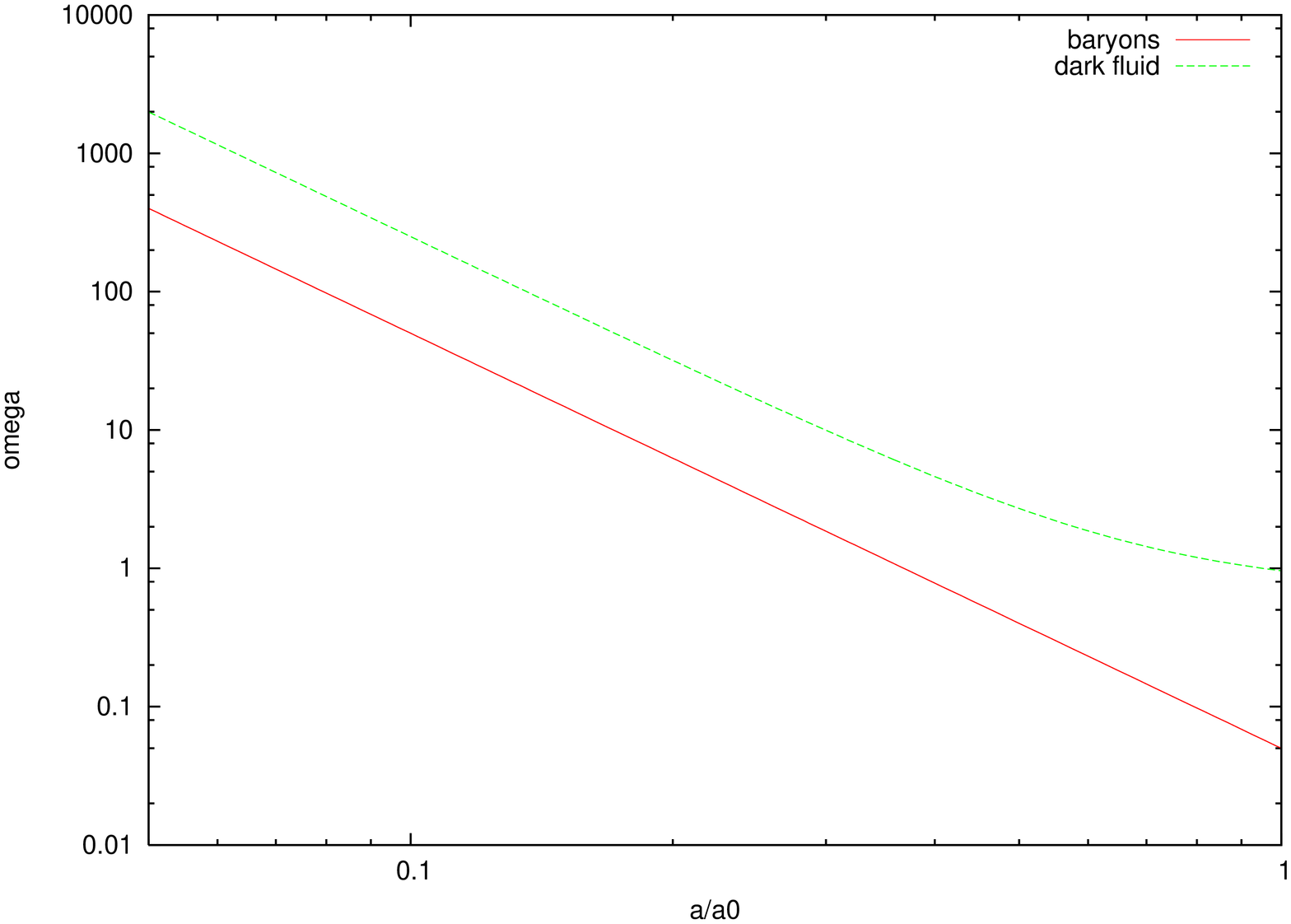}
\caption{Cosmological evolution of the density of the dark fluid scalar field in
comparison to the densities of baryonic matter and radiation.}
\label{cosmo}
\end{figure}

\noindent Modelling a dark fluid is not an easy task, as such a model should be able to describe simultaneously dark matter and dark energy properties. In the literature, only a few dark fluids have been investigated, the most well-known being the Chaplygin Gas (see Kamenshchik et al., 2001, Bilic et al., 2002 and R. Bean et al., 2003). In this paper, I consider a particular model using a scalar field, which was first introduced in (Arbey, 2006).\\
\\
Dark energy shares with vacuum energy the property of having a negative pressure and a quasi-constant energy density. Therefore, a standard scalar field is often referred as a good pretendant for dark energy, and models of dark energy based on scalar fields are generally referred as quintessence scalar fields (Peebles et al., 1988). An important question in these models is the choice of the potential of the scalar field, as it completely determines the behaviour of the field throughout the expansion of the Universe. Many potentials have already been investigated, and some of them, such as the decreasing exponential potential, have been already excluded by the cosmological data. However, it is still unclear how to choose the potential, and the main idea is to try to use potentials which can originate from physics motivated theories.\\
\\
The same problem appears when trying to build a dark fluid model, but the potential choice is now more constrained, as the scalar field has to behave like matter at local scale to account for observations in galaxies and clusters, and like dark energy at cosmological scales. The studies in (Arbey et al., 2001/2002/2003) have already shown that it is possible to describe dark matter in galaxies and at cosmological scales using a complex scalar field, provided the field has a mass term in the potential. We have shown in particular that with a model based on a complex scalar field $\phi$ with an internal rotation $\phi = \sigma e^{i\omega t}$ and based on the Lagrangian density:
\begin{equation}
{\cal L} \; = \;g^{\mu \nu} \, \partial_{\mu} \phi^{\dagger} \, \partial_{\nu} \phi\; - \; V \left( |\phi| \right) \;\; ,
\end{equation}
associated to a quadratic potential $V(\phi) =m^2 \phi^\dagger \phi$, the extended rotation curves of spiral galaxies can be reproduced, as can be seen for example in Fig.~\ref{gal}. To retain this dark matter behaviour, we will consider dark fluids demonstrating a $m^2 \phi^\dagger \phi$ behaviour in their potentials.\\
\\
Scalar fields have shown a great polyvalence, as they can lead to either a dark energy behaviour, or a matter behaviour. Therefore, it is quite natural to model a dark fluid using also a scalar field. However, an important question remains: how can the behaviour of the dark fluid be similar to dark matter at local scales and similar to dark energy at cosmological scales. Remarking that the density of dark fluid at cosmological scales today is of the order of the critical density, i.e. $\rho^0_c \approx -9\times-10^{-29}\mbox{ g.cm}^{-3}$, and that the average matter density in galaxies is of the order $\rho^{\mbox{gal}} \approx 5 \times 10^{-24}\mbox{ g.cm}^{-3}$, an answer can arise: the dark fluid needs to be inhomogeneous.\\
\\
In (Arbey, 2006), the dark fluid model using the potential
\begin{equation}
V(\phi)=m^2 |\phi|^2 + A e^{-B |\phi|^2} \label{potential}
\end{equation}
was investigated. It is formed with a quadratic term giving a mass to the field, and mostly responsible for the dark matter behaviour, and a decreasing exponential part, which can find an origin in some high energy theories and which can recall quintessence potentials, responsible for the dark energy behaviour. In the regions of spacetime where the scalar field density is large enough, for example around galaxies or in the Early Universe, the quadratic part of the potential would dominate, and where the density is small, the second part of the potential dominates, leading to a repulsive vacuum energy-like behaviour. Then, such a potential leads to a Universe highly inhomogeneous today.\\
It was also shown that such a potential can lead to the cosmological behaviour
illustrated in Fig. \ref{cosmo}, which is consistent with the observations, provided the different parameters are chosen adequately.
To fix the parameters, we consider three scales: the mass $m$ is fixed by galaxy
scales, confronting the model predictions to galaxy rotation curves; the $B$
parameters is fixed in order for the field to behave mostly as dark matter at typical cluster scales; and $A$ is chosen in order for the field to be consistent with the cosmological observations revealing a dark energy behaviour. We have then $m \sim 10^{-23}$ eV, $B \sim 10^{-22} \mbox{ eV}^{-2}$ and $A \sim \rho_0^{dark\, energy}$.
With this choice of parameters, we have shown that the scalar field dark fluid model is able to reproduce galaxy rotation curves, to condensate at cluster scales, and to have today a negative pressure at cosmological scales.\\
\\
The question of the choice of the potential remains crucial. We would like it to be physics motivated, and to originate if possible from high energy theories. To
investigate the behaviour of the scalar field from the quantum physics point of view, we used in (Arbey \& Mahmoudi, 2007) an effective quantum field theory approach to determine how quantum fluctuations would affects the potential of Eq.
(\ref{potential}). We have shown in this way that such a potential would be modified by quantum fluctuations, especially if we expect it to be somehow non-minimally coupled to matter fields. This result can be understood in two different manners: either such a potential is not resistant to quantum fluctuation and is not viable from the quantum theory point of view, or it is an effective potential which already includes the quantum corrections. In both points of view, the choice of the potential is uncertain. In this context, a new interesting potential is currently under investigation:
\begin{equation}
V(\Phi) = \alpha \, \mbox{cotanh}\left( \frac{\beta}{\Phi^\dagger \Phi} \right)\;\;.
\end{equation}
It has the advantage of finding roots in brane theories, and yet no big difference in behaviour is expected from the one determined by the potential of Eq. (\ref{potential}).

\section{Conctusion}
\noindent The standard model of cosmology assumes the existence of two unknown dark components, dark matter and dark energy. We have seen that it is possible to replace both components with a unique component, the dark fluid, and to remain consistent with the cosmological data. The properties of the dark fluid are already severely constrained because of its dual dark energy/dark matter properties, but it is nevertheless possible to model it with a simple and usual complex scalar field. The main question of the model is the choice of the scalar field potential, but an adequate choice can lead to a fluid explaning at the same time the dark matter observations at local scales, and the dark energy behaviour at cosmological scales. The relations between such a model and quantum theories are however not obvious, and still need to be worked out. However, to conclude, the dark fluid models are viable alternatives to the standard cosmological model, and need to be further investigated.


\begin{thebibliography}{}
\bibitem{}A. Arbey, astro--ph/0506732 (2005).

\bibitem{}A. Arbey, Phys. Rev. D 74 (2006), 043516.

\bibitem{}A. Arbey, The Open Astron. J. 1 (2008), 27 [arXiv: 0812.3122].

\bibitem{}A. Arbey, J. Lesgourgues \& P. Salati, Phys. Rev. D 64 (2001), 123528; Phys. Rev. D 65 (2002), 083514; Phys. Rev. D 68 (2003), 023511.

\bibitem{}A. Arbey \& F. Mahmoudi, Phys. Rev. D 75 (2007), 063513.

\bibitem{}R. Bean \& O. Dore, Phys.Rev. D 68 (2003), 023515.

\bibitem{}N. Bilic, G.B. Tupper \& R.O. Viollier, Phys. Lett. B 535 (2002), 17.

\bibitem{}A. Kamenshchik, U. Moschella \& V. Pasquier, Phys. Lett. B 511 (2001),
265.

\bibitem{}E. Komatsu et al., arXiv: 0803.0547 [astro-ph].

\bibitem{}P.J.E. Peebles \& B. Ratra, Astrophys. J. 325 (1988), L17.

\end{thebibliography}
\end{document}